\documentstyle[psfig,multicol,prl,aps]{revtex}

\begin{document}
\draft

\title{Non-Haldane Spin Liquid Models with Exact Ground States}

\author{A. K. Kolezhuk\cite{perm}}
\address{Institut f\"{u}r Theoretische Physik,
Universit\"{a}t Hannover, Appelstr. 2, 30167 Hannover, Germany\\
and Institute of Magnetism, National Academy of 
Sciences and Ministry of Education of Ukraine,
252142 Kiev, Ukraine}

\author{H.-J. Mikeska}
\address{Institut f\"{u}r Theoretische Physik,
Universit\"{a}t Hannover, Appelstr. 2, 30167 Hannover, Germany}

\date{\today}

\maketitle

\begin{abstract}
We present a family of spin ladder models which admit exact solution
for the ground state and exhibit non-Haldane spin liquid properties as
predicted recently by Nersesyan and Tsvelik [Phys. Rev. Lett. {\bf
78}, 3939 (1997)], and study their excitation spectrum using a simple
variational ansatz.  The elementary excitation is neither a magnon nor
a spinon, but a pair of propagating triplet or singlet solitons
connecting two spontaneously dimerized ground states.  Second-order
phase transitions separate this phase from the Haldane phase and the
rung-dimer phase.
\end{abstract}

\pacs{75.10.Jm, 75.40.Cx, 75.40.Gb, 75.40.-s}

\begin{multicols}{2}

It is well known that one-dimensional (1d) Heisenberg antiferromagnets
can exhibit several types of disordered ``quantum spin liquid''
phases. The spin-${1\over2}$ chain has a unique disordered gapless
ground state with power-law decay of spin correlations, and its
elementary excitations are pairs of spinons carrying spin ${1\over2}$
\cite{FaddeevTakht81}.  The ground state of the frustrated
$S={1\over2}$ chain with sufficiently strong next-nearest neighbor
interaction is doubly degenerate, the excitations are also spinon
pairs, but the spectrum is gapful
\cite{MajumdarGhosh69,ShastrySutherland81,Haldane82}; in presence of
any finite exchange alternation along the chain the spinon pairs get
confined into well defined magnon excitations \cite{Nakamura+}.  The
spin-$1$ (Haldane) chain has a unique spin-liquid ground state with a
gap above it formed by a triplet of magnons carrying spin $S=1$
\cite{Haldane83}.  The two-leg $S={1\over2}$ ladder, i.e., two
Heisenberg $S={1\over2}$ chains coupled by a transverse exchange, also
has a disordered gapful ground state with magnons as elementary
excitations \cite{DagottoRice96rev}, and is believed to be essentially
in the same phase as the Haldane chain, as well as frustrated
$S={1\over2}$ chain with alternating exchange
\cite{Hida92,White96,KM97}.

Recently, Nersesyan and Tsvelik \cite{NersesyanTsvelik97} have
proposed an interesting example of a 1d {\em ``non-Haldane
spin-liquid''\/} which has a gapped spectrum but whose excitations are
neither spinons nor magnons.  Using field-theoretical arguments, they
have shown that under certain conditions a two-leg $S={1\over2}$
Heisenberg ladder with additional leg-leg biquadratic interaction
enters spontaneously dimerized phase with the excitation spectrum
determined by the two-particle continuum, and identified the
elementary excitations as pairs of singlet and triplet domain walls
connecting the two dimerized ground states.

In this Letter we present a set of models which exhibit non-Haldane
spin liquid properties as predicted by Nersesyan and Tsvelik, and
whose ground state can be found {\em exactly.\/} We study their
excitation spectrum within a simple variational approach, and discuss
phase transitions into the Haldane and other phases.

We start from a more general ladder Hamiltonian which includes also
transverse interaction along the ladder diagonals and two additional
biquadratic interactions. The model
is described by the Hamiltonian
\FL
\begin{eqnarray} 
\label{ham}
&&\widehat{H}= \sum_{i} J ( {\mathbf S}_{1,i}{\mathbf S}_{1,i+1} +
	{\mathbf S}_{2,i}{\mathbf S}_{2,i+1} )
+ J_{r} {\mathbf S}_{1,i} {\mathbf S}_{2,i} \\
&&	+V ({\mathbf S}_{1,i}{\mathbf S}_{1,i+1}) 
	({\mathbf S}_{2,i}{\mathbf S}_{2,i+1})
+J_{d}({\mathbf S}_{1,i} {\mathbf S}_{2,i+1} 
	+{\mathbf S}_{2,i} {\mathbf S}_{1,i+1}) \nonumber\\ 
&&+ K\{({\mathbf S}_{1,i}{\mathbf S}_{2,i+1})
	({\mathbf S}_{2,i}{\mathbf S}_{1,i+1})
- ({\mathbf S}_{1,i}{\mathbf S}_{2,i}) 
	({\mathbf S}_{1,i+1}{\mathbf S}_{2,i+1}) \}\,,\nonumber
\end{eqnarray}  
where the indices $1$ and $2$ distinguish lower and upper legs, and
$i$ labels rungs.  The model considered by Nersesyan and Tsvelik
corresponds to $J_{d}=K=0$.  To construct the ground state $\Psi_{0}$ for the
Hamiltonian (\ref{ham}), we will use the technique of matrix-product
(MP) states \cite{Fannes+,Klumper+}.  We start from the following
ansatz:
\begin{eqnarray} 
\label{mpansatz} 
 \Psi_{0} &=& \text{tr}\big\{ g_{1}(\widetilde{u})\cdot
g_{2}(u)\cdots g_{2N-1}(\widetilde{u})\cdot
g_{2N}(u)\big\} \,,\\
 g_{i}(u)&=& 
u\cdot\widehat{\openone}|s\rangle_{i} 
+\sigma^{+1}|t_{+1}\rangle_{i}+\sigma^{-1}|t_{-1}\rangle_{i}
+\sigma^{0}|t_{0}\rangle_{i} \nonumber
\end{eqnarray}
Here $|s\rangle_{i}$ and $|t_{\mu}\rangle_{i}$ are the singlet and
triplet states of the $i$-th rung, $2N$ is the total number of rungs
(periodic boundary conditions are assumed),
$\widehat{\openone}$ is the 2$\times$2 unit matrix, $\sigma^{\mu}$ are
the Pauli matrices, and $u$, $\widetilde{u}$ are free parameters. 
A simpler version of this ansatz with  $u=\widetilde{u}$ 
describes several known examples of valence bond type (VBS) states,
e.g., at $u=0$ the wave function $\Psi_{0}$ is the ground state of the
effective Affleck-Kennedy-Lieb-Tasaki (AKLT) chain \cite{AKLT} whose
$S=1$ spins are composed from pairs of $S={1\over2}$ spins of the
ladder rungs,
\begin{equation}
\label{S1biquad}
\widehat{H}=\sum_{n} {\mathbf
S}_{n} {\mathbf S}_{n+1}
-\beta ({\mathbf S}_{n}\cdot {\mathbf S}_{n+1})^{2},
\end{equation}
at $\beta=-{1\over3}$, and for $u=1$ or $u=\infty$ one obtains
two degenerate ground states of the Majumdar-Ghosh model
\cite{MajumdarGhosh69}. Originally (\ref{mpansatz}) with $u=\widetilde{u}$
was proposed in Ref.\ \onlinecite{Brehmer+96} as a variational wave
function, and recently it was used by us \cite{KM97} to construct
another class of exact ground states for a more general ladder model.
In the following, we set $u\not=\widetilde{u}$, then the state
$\Psi_{0}$ is dimerized and the translation for one rung leads to a
different state with the same energy.  The ansatz (\ref{mpansatz})
obeys rotational symmetry, i.e., $\Psi_{0}$ is a global singlet
\cite{Brehmer+96,KMY97}.

The Hamiltonian (\ref{ham}) can be  rewritten as a sum of
identical local terms coupling neighboring rungs,
$\widehat{H}=\sum_{i}(\widehat{h}_{i,i+1}-E_{0})$. Let us demand that
the wave function (\ref{mpansatz}) is a zero-energy ground state of
$\widehat{H}$ (which can always be achieved by the appropriate choice
of $E_{0}$), then the following requirements have
to be fulfilled \cite{Klumper+}: (i) the local Hamiltonian
$\widehat{h}_{i,i+1}$ has to annihilate $\Psi_{0}$, which, due to the
product property of (\ref{mpansatz}), means that all elements of the
two matrix products $g_{i}(u) g_{i+1}(\widetilde{u})$,
$g_{i}(\widetilde{u})g_{i+1}(u)$ should be zero-energy eigenstates of
$\widehat{h}_{i,i+1}$; (ii) the other eigenstates of
$\widehat{h}_{i,i+1}$ should have positive energy. 
 Those two conditions fix the structure of the local Hamiltonian
as follows:
\begin{equation}
\label{ham-proj}
\widehat{h}_{i,i+1}=
\sum_{J=0,1,2}\sum_{M=-J}^{J}\lambda_{J}|\psi_{JM}\rangle\langle
\psi_{JM}| \,,
\end{equation}
where the eigenvalues $\lambda_{J}>0$, and $|\psi_{JM}\rangle$ are the
components of the positive-energy multiplets constructed from the states of
the four-spin plaquette $(i,i+1)$:
\begin{eqnarray} 
\label{lgs} 
&&|\psi_{00}\rangle=\big[3+(u\widetilde{u})^{2}\big]^{-1/2}
\big\{\sqrt{3}|ss\rangle
+u\widetilde{u}\,|tt\rangle_{J=0}\big\},\nonumber\\
&&|\psi_{1}\rangle=[2+f^{2}]^{-1/2}\big\{ 
f|tt\rangle_{J=1}+|st\rangle+|ts\rangle\big\} \\
&&|\psi_{2}\rangle=|tt\rangle_{J=2},
\quad f\equiv (u+\widetilde{u})/\sqrt{2}.\nonumber 
\end{eqnarray}
Here we use the notation  $|tt\rangle_{J=1}$ for the triplet of
states with the total spin $J=1$ constructed, in turn, from two
triplets on rungs $i$ and $i+1$, etc.

\begin{figure}
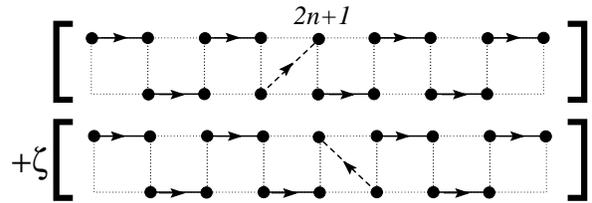

\caption{\label{fig:ntexc} The states $|n\rangle_{t,s}^{\zeta}$ used
in Eq.\ (\protect\ref{soliton}), in a special case of the model
(\protect\ref{generic}).  Thick solid lines indicate singlet bonds,
and thick dashed lines can be either singlets or triplets. Arrows
indicate the ``direction'' of the singlet bonds [i.e.,
$|s_{1\rightarrow2}\rangle=2^{-1/2}(|\uparrow_{1}\downarrow_{2}\rangle
-|\downarrow_{1}\uparrow_{2}\rangle)$].}
\end{figure}

\begin{figure}
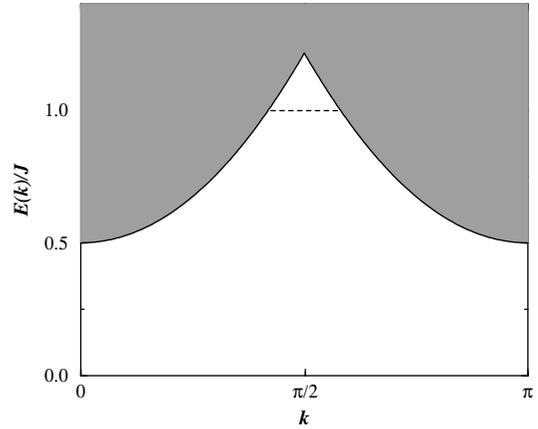

\caption{\label{fig:ntspec}
The excitation spectrum of the model
(\protect\ref{generic}). The continuum is
determined by free two-soliton states, its lowest boundary is 16-fold
degenerate. The dashed line is determined
by the Haldane triplet excitation (\protect\ref{Haldane}) and
indicates a variational estimate for bound soliton-antisoliton states.}
\end{figure}

Now we demand that the structure (\ref{ham-proj}) is compatible with
the desired form of the Hamiltonian (\ref{ham}), which yields the
connection between the parameters $J$, $J_{r}$, $J_{d}$, $V$, $K$ on
one hand, and the local eigenvalues $\lambda_{J}$ and singlet weight parameters
$u$, $\widetilde{u}$ of the ground state wave function on the
other. Those solutions can be classified into the following three types:

{\em (A)  ``Checkerboard-dimer'' model} with $K=0$, $J_{d}\not=0$:
\begin{eqnarray} 
\label{familyA}
&& u = \pm1, \quad  \widetilde{u} =\mp1,\quad  V=4J/3,\quad K=0,\\
&& \lambda_{1}=1,\quad \lambda_{0}=3x/8, \quad 
\lambda_{2}=3(1-x),\quad 0\leq x \leq 1,\nonumber\\
&&J_{r}=(8J/3)(2-3x)/(4-3x),\quad
 J_{d}=J_{r}/2,\quad J>0,\nonumber 
\end{eqnarray}
the ground state energy density per rung is $E_{0}=-{3\over4}J$, and
$x$ is an arbitrary parameter.  Two degenerate ground states are
simply checkerboard-type products of singlet bonds along the ladder
legs.  A generic example from this family is the model at
$x={2\over3}$ with purely biquadratic interchain interaction:
\begin{equation}
\label{generic} 
J_{r}=J_{d}=K=0, \quad J=3V/4>0\,.
\end{equation} 
\mbox{\psfig{figure=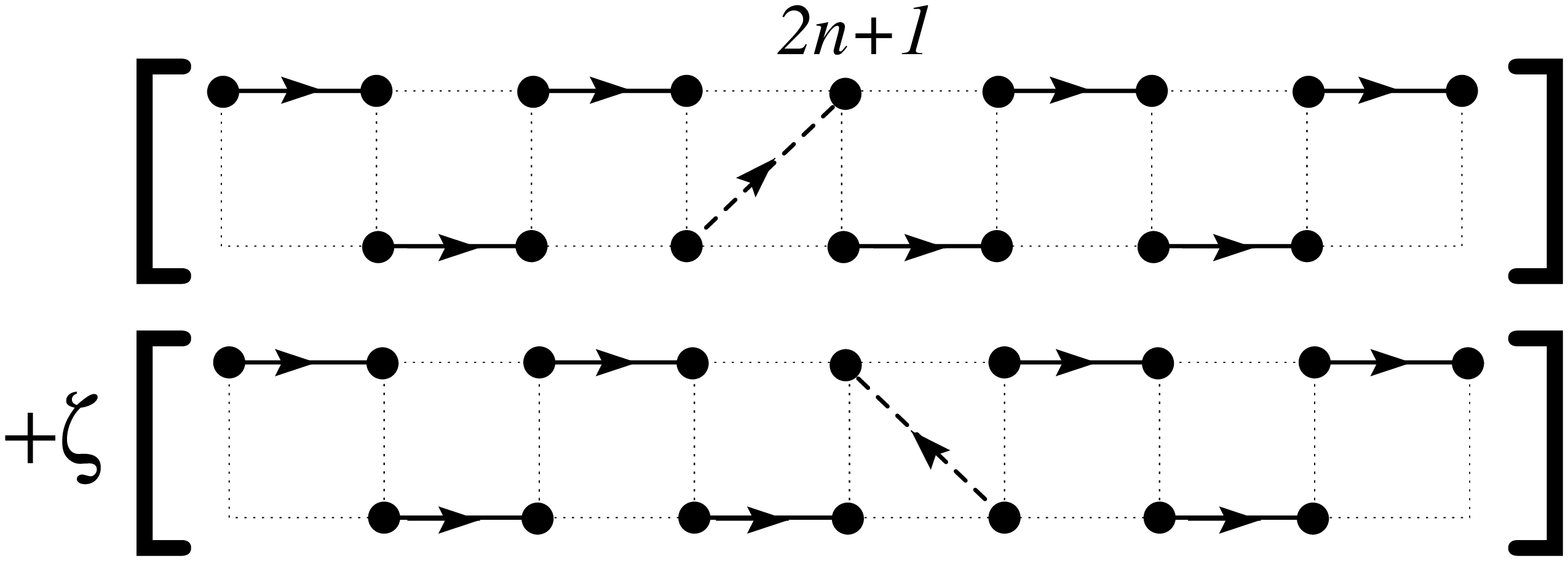,width=80mm,angle=-90.}}
  \vskip 2mm\nopagebreak
\noindent\parbox[t]{3.40in} {\protect\small
   FIG. \ref{fig:ntexc}
The excited states $|n\rangle_{t,s}^{\zeta}$  of the model
(\protect\ref{familyA}), used in Eq.\ (\protect\ref{soliton}).  
 Thick solid lines indicate singlet bonds, and
thick dashed lines can be either singlets or triplets. Arrows indicate the
``direction'' of the singlet bonds
[i.e., $|s_{1\rightarrow2}\rangle=2^{-1/2}(|\uparrow_{1}\downarrow_{2}\rangle 
-|\downarrow_{1}\uparrow_{2}\rangle)$].}
\vskip -5mm
\mbox{\psfig{figure=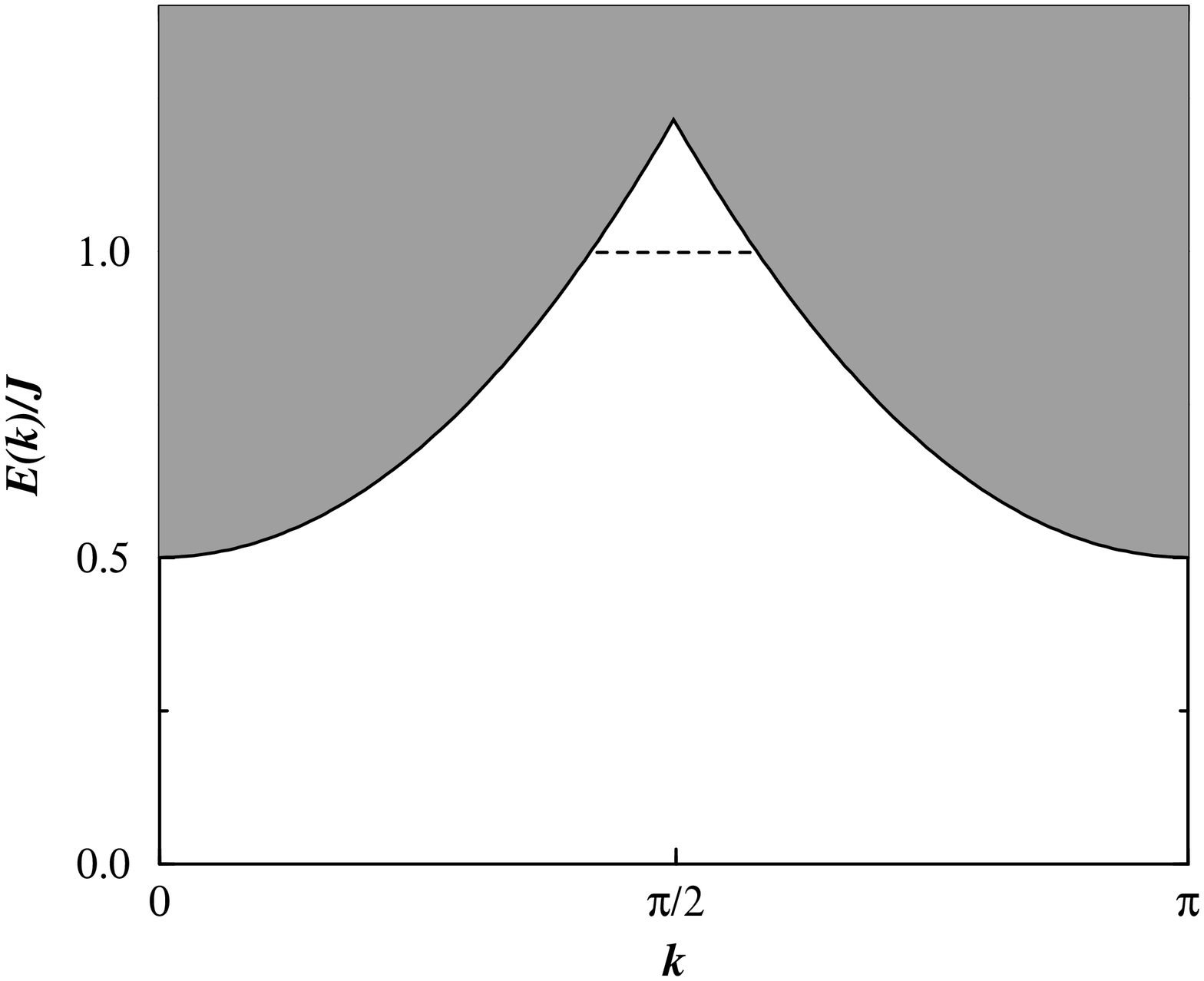,width=80mm,angle=-90.}}
  \vskip 2mm\nopagebreak
\noindent\parbox[t]{3.40in} {\protect\small
   FIG. \ref{fig:ntspec}
The excitation spectrum of the model
(\protect\ref{generic}). The continuum is
determined by free two-soliton states, its lowest boundary is 16-fold
degenerate. The dashed line is determined
by the Haldane triplet excitation (\protect\ref{Haldane}) and
indicates a variational estimate for bound soliton-antisoliton states.}
\vskip 4mm
\noindent
This ``generic'' model lies within the class of Hamiltonians
considered by Nersesyan and Tsvelik.  At $x=1$ the eigenvalue
$\lambda_{2}$ vanishes, indicating a first-order transition into the
fully polarized ferromagnetic state.

{\em (B) Multicritical model:}
\begin{eqnarray} 
\label{familyB}
&& u=-\widetilde{u},\quad \lambda_{0}=0,\quad
\lambda_{2}=3\lambda_{1},\nonumber\\
&& J_{r}=V=4J/3, \quad J_{d}=J_{r}/2,\quad
 K=0,\quad J>0\,.
\end{eqnarray}
This model has a remarkable property: {\em any} wave function
$\Psi_{0}(u)$ of the form (\ref{mpansatz}) with $u=-\widetilde{u}$ is
a ground state with the same energy per rung
$E_{0}=-{3\over4}J$. One can show that 
two ground state wave functions with different values of $u$ are
{\em asymptotically orthogonal\/} in thermodynamic limit $N\to\infty$:
$\langle \Psi_{0}(u)|
\Psi_{0}(u')\rangle=z^{N}$, $z(u,u')\leq 1$,  
so that the degeneracy of the ground state is exponentially large.
It is easy to observe that the model
(\ref{familyB}) is  a particular case of (\ref{familyA}) at $x=0$,
so that the model (\ref{familyA}) has another phase transition point at
$x=0$; below we will argue that this transition is of the first order.

{\em (C) Model with two second-order phase boundaries} with
 $J_{d}=0$, $K\not=0$:
\begin{eqnarray} 
\label{familyC} 
&& u=-\widetilde{u},\;\;
K=J_{r}=\lambda_{0}(u^{2}-1)(u^{2}+3)/2,
\;\; J_{d}=0,\\
&&
 V=\lambda_{0}(5u^{4}+2u^{2}+9)/4,\;\;
 J=3\lambda_{0}(u^{4}+10u^{2}+5)/16,
\nonumber\\
&&\lambda_{1}=\lambda_{0}(3u^{4}\!+ 14u^{2}\!+15)/8,\;
\lambda_{2}=\lambda_{0}(5u^{4}\!+18u^{2}\!+9)/8,\nonumber
\end{eqnarray}
the g.s.\ energy per rung is
$E_{0}=-{3\over64}\lambda_{0}(7u^{4}+22u^{2}+19)$. This is a
one-parametric family of models since $u$ is arbitrary (the parameter
$\lambda_{0}$ just sets the energy scale and thus is irrelevant).  A
particular case $u=\pm1$ again leads to the ``generic'' model
(\ref{generic}).  One can readily observe that at $u=0$ or $u=\infty$
the ground state is no more dimerized.  The state with $u=0$ describes
the ground state of an effective $S=1$ chain (\ref{S1biquad}) with
$\beta=-{1\over3}$; the state with $u=\infty$ corresponds to a product
of singlet bonds on the rungs. It is easy to calculate spin-spin and
dimer-dimer correlation functions $C_{S}(n)=\langle S^{z}_{1,i}
S^{z}_{1,i+n}\rangle$ and $C_{D}(n)=\langle D_{i}D_{i+n}\rangle$, here
$D_{i}={\mathbf S}_{1,i}\cdot({\mathbf S}_{1,i+1}-{\mathbf S}_{1,i-1})$:
\begin{eqnarray} 
\label{correl} 
C_{S}(n)&=&(u^{2}+3)^{-1}
(z_{+}z_{-})^{n} 
(\delta_{n,2k}-z_{-}\delta_{n,2k+1}), \\
C_{D}(n)&=&144u^{2}/(u^{2}+3)^{4},\quad 
z_{\pm}=(u\pm1)^{2}/(u^{2}+3)\,.\nonumber
\end{eqnarray}
One can see that the dimer correlations exhibit long-range order
vanishing for $u\to 0,\infty$, but remarkably there is no exponential
tail. The spin correlation length goes through zero at $u=1$ and
diverges at $u\to\infty$, however, there is no long-range spin order
at $u\to\infty$ since the amplitude of spin correlations 
vanishes in this limit.  Thus, one can conclude that the model
{\em (C)} exhibits two second-order phase transitions: into the
Haldane phase at $u=0$ and into the rung-dimer phase at $u=\infty$. We
will show below that those transitions are characterized by vanishing
singlet and triplet gaps, respectively.

By induction with respect to the ladder length one can prove that in
cases {\em (A)} and {\em (C)} the two ground states given by the MP
ansatz are the only ground states of the system.

Elementary excitations of the model {\em (A)} can be easily
visualized as singlet or triplet diagonal bonds separating the two
ground states and thus being solitons in the dimer order (see Fig.\
\ref{fig:ntexc}). Since solitons can be created only in pairs, the
excitation spectrum is a two-particle continuum.  To study the
scattering soliton states, one may consider the ladder with $2N+1$
rungs and periodic boundary conditions, and write down a simple
single-soliton variational state with certain value of momentum $p$
and parity $\zeta=\pm1$:
\begin{equation} 
\label{soliton}
|p\rangle_{t,s}^{\zeta}=\sum_{n} 
e^{ip(2n+1)}|n\rangle_{t,s}^{\zeta} \,,
\end{equation}
Here the momenta are defined in terms of the Brillouin zone of
non-dimerized ladder, so that $p\in[0,\pi]$.  The states
$|n\rangle_{t,s}^{\zeta}$ are shown in Fig.\ \ref{fig:ntexc}; in a MP
formulation they can be written as
\begin{eqnarray} 
\label{genexc}
&&|n\rangle_{s,t}^{\zeta}=\prod_{i=1}^{n}
\Big(g_{2i-1}(\widetilde{u})\, g_{2i}(u)\Big)\,  g_{2n+1}^{(s,t)}
\!\prod_{i=n+1}^{N} \! g_{2i}(\widetilde{u})\,  g_{2i+1}(u)
\,,\nonumber\\
&&g^{s}_{\zeta}= g(u) -\zeta g(\widetilde{u}),\quad
g^{t}_{\zeta,\mu}= \sigma^{\mu} g(u) +\zeta  g(\widetilde{u})\sigma^{\mu}\,.
\end{eqnarray}
{}
\vskip -15mm
\mbox{\hspace*{-6mm}\psfig{figure=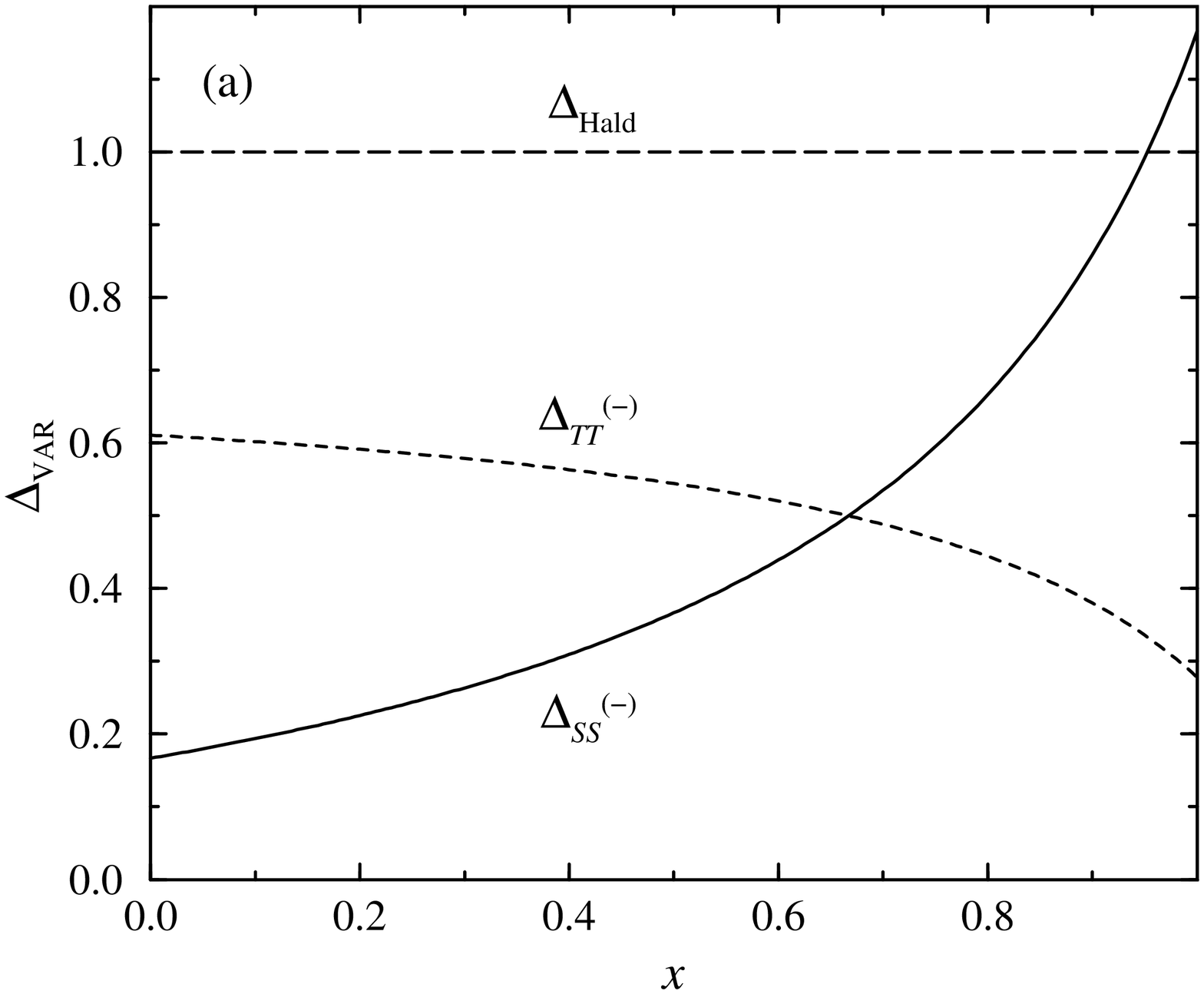,width=90mm,angle=-90}}

\begin{figure}
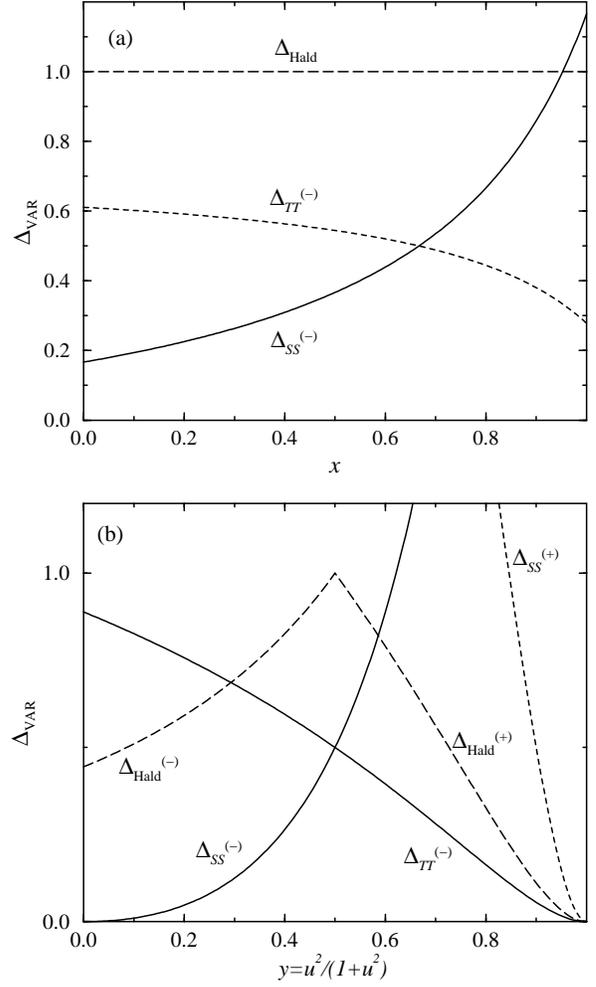

\caption{\label{fig:ntgaps}
(a) The gaps of different variational excitations for the model
(\protect\ref{familyA}); (b) the same for model
(\protect\ref{familyC}). Here $\Delta_{SS}^{\zeta}$, $\Delta_{TT}^{\zeta}$,
and $\Delta_{H}^{\zeta}$ denote the gaps of singlet-singlet,
triplet-triplet soliton pairs, and the Haldane triplet, respectively,
and $\zeta=\pm1$ is the parity.
}
\end{figure}

\vskip -5mm
\mbox{\hspace*{-6mm}\psfig{figure=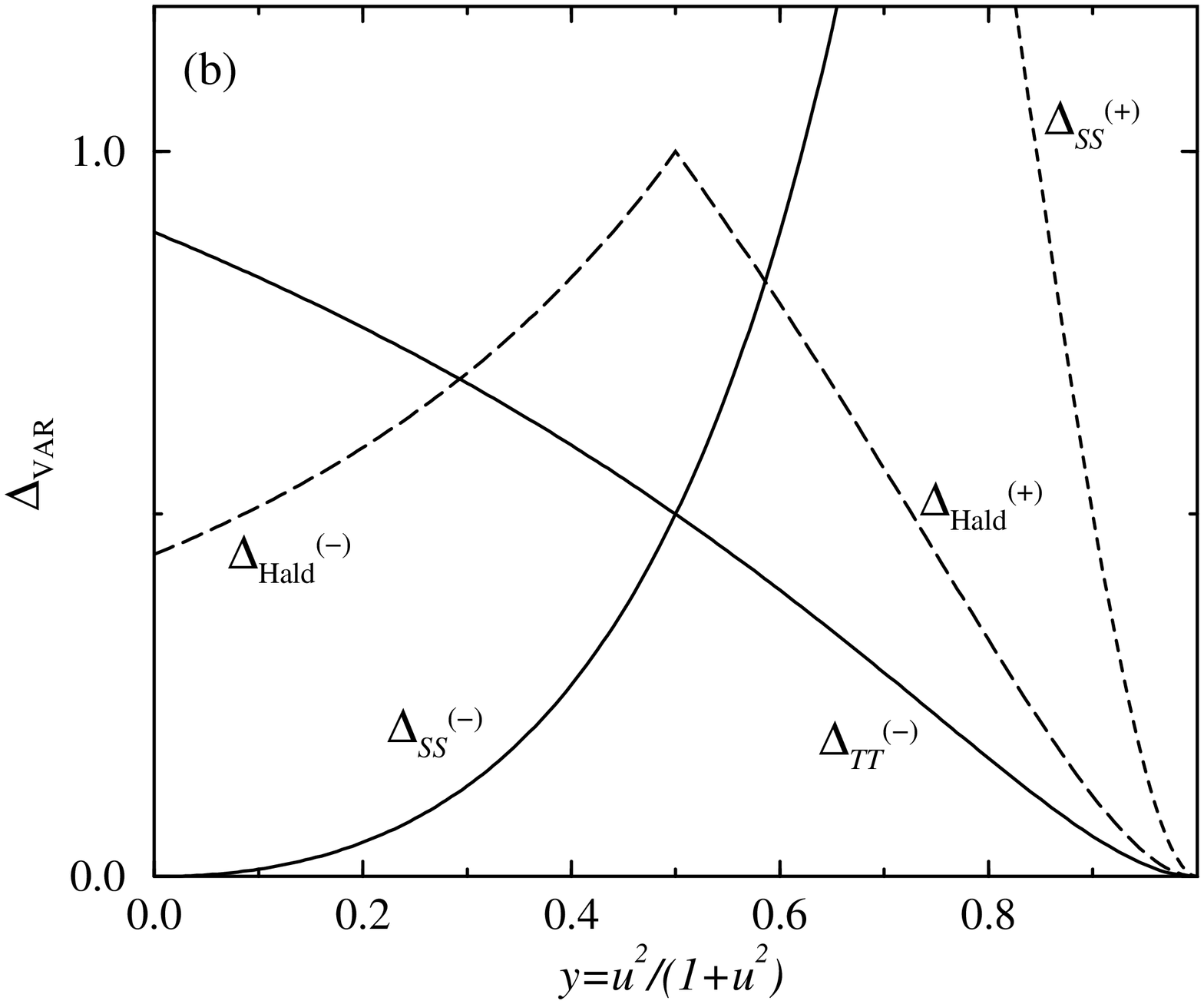,width=90mm,angle=-90}}
 \vskip 2mm\nopagebreak
\noindent\parbox[t]{3.40in} {\protect\small
   FIG. \ref{fig:ntgaps}
(a) The gaps of different variational excitations for the model
(\protect\ref{familyA}); (b) the same for model
(\protect\ref{familyC}). Here $\Delta_{SS}^{\zeta}$, $\Delta_{TT}^{\zeta}$,
and $\Delta_{H}^{\zeta}$ denote the gaps of singlet-singlet,
triplet-triplet soliton pairs, and the Haldane triplet, respectively,
and $\zeta=\pm1$ is the parity.
}
\vskip 2mm
\noindent
Here $\mu=0,\pm1$ denotes the $z$-projection of spin of the triplet
excitation.  Another candidate for the role of the elementary
excitation is a magnon (the Haldane triplet); the corresponding
variational wavefunction can be again written in the form
(\ref{soliton}) with
\begin{eqnarray} 
\label{Haldane} 
&&|n\rangle_{H}^{\zeta}
=\prod_{i=1}^{n-1}
\Big(g_{2i-1}(\widetilde{u})\, g_{2i}(u)\Big)\,  g^{H}_{\zeta}
\!\prod_{i=n+1}^{N} \! g_{2i-1}(\widetilde{u})\,  g_{2i}(u)
\,,\nonumber\\
&&g^{H}_{\zeta,\mu}= \sigma^{\mu} g_{2n-1}(\widetilde{u}) g_{2n}(u) 
+ \zeta  g_{2n-1}(\widetilde{u})\sigma^{\mu} g_{2n}(u)\,.
\end{eqnarray}
The variational dispersion laws have the following form:
\begin{eqnarray} 
\label{gendisp}
&& \varepsilon (p)=e_{0}/[1+2c_{0}A(z,p)], 
\nonumber\\
&& A(z,p)=(\cos(2p)-z)/(1+z^{2}-2z\cos(2p)),
\end{eqnarray}
and for the 
model (\ref{familyA}) one has, in $J$ units,
\begin{eqnarray} 
\label{resA} 
&&c_{0}^{s,t}=z_{s,t}(1/2-\zeta),\;\;
z_{s,t}=1/4,\;\; z_{H}\!=c_{0}^{H}\!=0,\;\; e_{0}^{H}\!=1\,,\nonumber\\
&&e_{0}^{s}={4+3x\over (4-2\zeta)(4-3x)},\quad
e_{0}^{t}={44-39x\over 6(2-\zeta)(4-3x)}\,.
\end{eqnarray}
One can see that the lowest energy is always reached for the
odd-parity states ($\zeta=-1$). The Haldane triplet is in this case
dispersionless, and has a high energy equal to 1.  The elementary
excitation is a soliton-antisoliton pair, and for the scattering
states its energy is given by
\begin{equation} 
\label{twop} 
\widetilde{E}(k,q)=\varepsilon_{s,t}\big[ (k+q)/2\big]
+\varepsilon_{s,t}\big[ (k-q)/2\big]\,,
\end{equation}  
where $k$ and $q$ are the total and relative momentum.  For
$x={2\over3}$ [i.e., for the ``generic'' model (\ref{generic}) with
zero transverse exchange] the energies of triplet and singlet solitons
coincide.  The lowest boundary $E(k)$ of the continuum described by
(\ref{twop}) at $x={2\over3}$ is shown in Fig.\ \ref{fig:ntspec}.  The
gap is given by $E(0)=E(\pi)={1\over2}J$, and the lowest excitation
has a 16-fold degeneracy because the states of a soliton pair can be
classified into two singlets $(ss)$ and $(tt)_{J=0}$, three triplets
$(st)$, $(ts)$, and $(tt)_{J=1}$ and one quintuplet $(tt)_{J=2}$. The
energy of the Haldane triplet is lower than the continuum boundary in
the vicinity of the zone center $k={\pi\over2}$, indicating possible
presence of bound soliton-antisoliton states.  If the transverse
exchange is switched on (i.e., $x\not={2\over3}$), the singlet-triplet
degeneracy is lifted, and for $x<{2\over3}$ ($x>{2\over3}$) the lowest
excitation is determined by singlet (triplet) solitons,
respectively. Behavior of the corresponding gaps is shown in Fig.\
\ref{fig:ntgaps}(a); one can see that for both phase transition points
$x=0$ and $x=1$ the gaps remain finite, which suggests that the
transition to the ``multicritical'' state at $x=0$ is of the first
order.

The ans\"atze (\ref{genexc}), (\ref{Haldane}) can be used for the
model {\em (C)} as well. One again obtains the dispersion laws
of the form (\ref{gendisp}), with
\begin{eqnarray} 
\label{resC} 
&& c_{0}^{s}=z_{s}(1+\zeta z_{s}^{-1/2})/2,\;\;
c_{0}^{t}=z_{t}(1-\zeta z_{t}^{-1/2})/2\nonumber\\
&& e_{0}^{s}=
12 u^{2}/\big\{(u^{2}+3)^{2}(1+\zeta z_{s}^{1/2})\big\}\,,\\
&& e_{0}^{t}=
4(u^{2}+2)/\big\{(u^{2}+3)^{2}(1-\zeta z_{t}^{1/2})\big\}\,,\nonumber \\
&& c_{0}^{H}=z_{H}
   + {\zeta z_{H}^{1/2} (1-z_{H}) \over 2(1+\zeta z_{H}^{1/2})}\,,\;\;
 e_{0}^{H}={8 z_{t}^{1/2}\over (u^{2}+3)(1+\zeta
z_{H}^{1/2})}\,,\nonumber\\
&& z_{s}^{1/2}={u^{2}-3\over u^{2}+3},\quad
z_{t}^{1/2}={u^{2}+1\over u^{2}+3},\quad
z_{H}^{1/2}={u^{2}-1\over u^{2}+3}.\nonumber
\end{eqnarray}
Behavior of the gaps is shown in Fig.\ \ref{fig:ntgaps}(b) as a
function of parameter $y=u^{2}/(1+u^{2})$.  Again, the lowest
excitations are always soliton pairs.  At $y\to0$ the
odd-singlet soliton gap goes to zero, indicating the second-order
transition to the Haldane phase.  At $y\to1$ three gaps (of
even-singlet and odd-triplet solitons and of the even Haldane triplet)
vanish, signaling another second-order transition into the rung-dimer
phase. Actually, it follows from (\ref{resC}) that at $y\to0$ $(1)$ the
whole continuum of singlet (triplet) soliton pairs collapses to zero.

{\em Acknowledgements.\/} A.K.\ gratefully acknowledges the
hospitality of Hannover Institute for Theoretical Physics.  This work
was supported in part by the German Ministry for Research and
Technology (BMBF) under the contract 03MI4HAN8 and by the Ukrainian
Ministry of Science (grant 2.4/27).

\end{multicols}
\newpage

\end{document}